# MODALITY-INDEPENDENT EXPLAINABLE DETECTION OF INACCURATE ORGAN SEGMENTATIONS USING DENOISING AUTOENCODERS

*Levente Lippenszky, István Megyeri, Krisztian Koos, Zsófia Karancsi, Borbála Deák-Karancsi, András Frontó, Árpád Makk, Attila Rádics, Erhan Bas and László Ruskó*
GE HealthCare

## ABSTRACT

In radiation therapy planning, inaccurate segmentations of organs at risk can result in suboptimal treatment delivery, if left undetected by the clinician. To address this challenge, we developed a denoising autoencoder-based method to detect inaccurate organ segmentations. We applied noise to ground truth organ segmentations, and the autoencoders were tasked to denoise them. Through the application of our method to organ segmentations generated on both MR and CT scans, we demonstrated that the method is independent of imaging modality. By providing reconstructions, our method offers visual information about inaccurate regions of the organ segmentations, leading to more explainable detection of suboptimal segmentations. We compared our method to existing approaches in the literature and demonstrated that it achieved superior performance for the majority of organs.

## 1. INTRODUCTION

Inaccurate segmentation of organs at risk (OARs) can lead to errors in dose calculation during radiation therapy planning if left undetected by the clinical user. This challenge is further complicated by automation bias, where clinicians tend to overly rely on auto-segmentations [1]. Such inaccuracies can result in suboptimal treatment delivery and an increased risk of adverse effects on normal tissues. Therefore, it is crucial to minimize OAR segmentation errors for treatments designed to avoid OARs safely [2].

Some of the previous studies utilized the Mahalanobis distance in the feature space of deep learning-based segmentation models to detect out-of-distribution (OOD) segmentations. Gonzalez et al. utilized the Mahalanobis distance on features extracted from the encoder of nnU-Net for OOD detection in lung lesions segmentation [3]. Woodland et al[1]. detected OOD liver segmentations by applying the Mahalanobis distance to the bottleneck features of a Swin UNETR model after reducing the dimensions using principal component analysis [4]. While useful, the application of these methods requires access to the segmentation models, and consequently, they are dependent on the imaging modality.

Other methods leveraged a statistical approach to detect OAR segmentation errors. Both Hui et al. and Altman et al. estimated the distributions of manually engineered features—such as shape, size characteristics of the segmentations, image-based features—and detected errors when these features deviated from their distributions [2], [5]. While these methods do not require access to the segmentation models, they are still dependent on the imaging modality due to the use of image intensity-based features. Furthermore, all the approaches output an inaccuracy score, typically without providing information to explain the assessment by the quality assurance system.

Sandfort et al. developed three distinct variational autoencoders (VAEs) to detect incorrect liver, spleen and kidney segmentations on CT scans. Each VAE was trained on auto-segmentations produced by a U-Net model specific to the corresponding organ [6]. Although modality-independence and explainability are applicable to this study, these aspects were not investigated by the authors.

To address these limitations, we developed an explainable and modality-independent method to detect inaccurate organ segmentations using denoising autoencoders (DAEs). In contrast to [6], our DAEs are designed to process multiple organs simultaneously and leverage information from neighboring organs. During training, we applied noise to ground truth organ segmentations, and the autoencoders were tasked to denoise them. For an organ auto-segmentation, we defined the inaccuracy score as the Dice loss between the auto-segmentation mask and the reconstructed mask produced by the model. As our method only requires binary segmentation masks, it is independent of the segmentation models and the imaging modality. We demonstrate modality independence through the application of our method to organ segmentations in the pelvis anatomy on MR scans and kidney segmentations on CT scans. The CT kidney use case contains both kidneys, while the MR pelvis use case covers seven organs including bladder, left and right femoral heads, penile bulb, prostate, rectum and urethra. By providing reconstructed masks, our method presents visual information

---

about inaccurate regions of organ auto-segmentations. This leads to improved explainability of the inaccuracy scores, which may facilitate better clinician-AI interaction and build trust in practical applications. To compare our method with existing solutions, we implemented a statistical approach following [2], [5] and two VAE-based approaches, one of which is described in [6].

## 2. METHODS

### 2.1. MR Pelvis Segmentation Models

To generate organ auto-segmentations for the MR pelvis use case, we utilized deep learning segmentation models described in Czipczer et al. [7]. Their method incorporated a localization module that employed 2D U-Net [8] segmentation models on axial, coronal and sagittal slices to find the center of the 3D bounding box for each OAR. Subsequently, a 3D U-Net was utilized to segment the OAR within the bounding box. The models were trained and evaluated on T2-weighted MR images.

### 2.2. CT Kidney Segmentation Models

We generated kidney auto-segmentations using Auto Segmentation, a deep learning-based solution for CT scans that was 510(k) cleared by the U.S. FDA and CE Marked [9]. The solution includes a localization model and 15 segmentation models covering 40 OARs. The localization model, utilizing the Inception-v3 architecture [10], is a slice-by-slice image classification model that categorizes CT slices into eight anatomical regions. Based on the localization results, 2D and 3D U-Net-like segmentation models are inferred on the respective slices.

### 2.3. Data

For the MR pelvis use case, training set consisted of 72 male cases with manually annotated ground truth segmentations for bladder, left and right femoral heads, penile bulb, prostate, rectum and urethra. Validation set contained 18 male cases. To increase the dataset size, each case was flipped over the mid-sagittal plane as pelvis is left-right symmetric with respect to the organs of interest. We utilized a test set containing 45 male cases. To increase the test set size, we rotated the scans along the x-, y- and z-axis by either 10 or -10 degrees using linear interpolation, which resulted in 270 scans in total. For evaluation, we generated auto-segmentations using the MR pelvis segmentation models. The corresponding ground truth segmentations were rotated accordingly using nearest neighbor interpolation. All the MR scans were collected privately.

For the CT kidney use case, training and validation sets comprised of 69 and 8 cases, respectively. Ground truth segmentations for left and right kidneys were annotated by medical professionals. The test set contained auto-segmentations generated by the CT kidney segmentation models on 157 CT scans. The scans in this use case originated from two data sources. 20 scans came from the CT-ORG public dataset [11], the rest were collected privately (see Section 5 for further details).

### 2.4. Proposed Method

We propose an algorithm to detect inaccurate organ segmentations utilizing denoising autoencoders. For each use case, we trained a multi-organ convolutional DAE using only ground truth segmentations (Figure 1). During training, we resampled each organ's ground truth segmentation to a common spacing and padded it to a common spatial size (Table 1). Then, we created the target tensor by stacking these processed segmentations channel-wise. Subsequently, we created the input tensor by applying noise to each channel of the target tensor. Noise generation involved either adding or removing random binary patches from the organ target segmentation, each with probability of 0.5. There were four hyperparameters controlling the noise generation: maximum number of patches, minimum and maximum patch size, and the sampling method for patch center. We sampled the center of a random patch either from within the organ target segmentation's foreground or from its bounding box. The former approach was applied to small organs, such as the urethra, which sparsely occupy their bounding boxes. For each organ, hyperparameters were optimized manually such that the signed Dice coefficients of the noisy input masks cover the [-1,1] interval roughly uniformly in the training set. We defined the signed Dice coefficient as $\text{sgn}(|I| - |T|)\frac{2|I \cap T|}{|I|+|T|}$, where $I$ and $T$ are the input and target binary segmentation masks, respectively, $\text{sgn}(\cdot)$ is the sign function and $|\cdot|$ denotes the count of 1s. We trained a 3D U-Net with residual units [12], as implemented in MONAI 1.3.0, to denoise the corrupted input segmentations, hyperparameters are summarized in Table 2. Average Dice loss across channels was used as the loss function, which was defined for an example $i$ with $C$ channels as $\frac{1}{C}\sum_{j=1}^{C}\left(1 - \frac{2|R_{i,j} \cap T_{i,j}|}{|R_{i,j}|+|T_{i,j}|}\right)$, where $R$ denotes the reconstructed segmentation after the forward pass. Optimization was performed using Adam [13] with a learning rate of $10^{-3}$. Models were trained until convergence and weights were saved at the lowest validation loss. Trainings were performed on an NVIDIA GeForce RTX 3090 GPU card with 24 GB memory.

During inference, organ auto-segmentations were preprocessed similarly as in training. Then, the auto-segmentations were stacked channel-wise and were inputted to the trained model. Inaccuracy score was defined as the Dice loss between the preprocessed auto-segmentation and the reconstruction (Figure 2).

| Hyperparameter | MR Pelvis | CT Kidney |
|---|---|---|
| Spacing (mm) | (1.5, 1.5, 1.5) | (1.0, 1.0, 3.0) |
| Spatial size | (336, 336, 240) | (700, 700, 620) |
| Bounding box size | - | (272, 160, 80) |

**Table 1.** Spacing and spatial size for the two use cases. Due to GPU memory constraints, tensors were cropped to a smaller spatial size using bounding box centered around the center of foreground mass for the CT kidney use case.

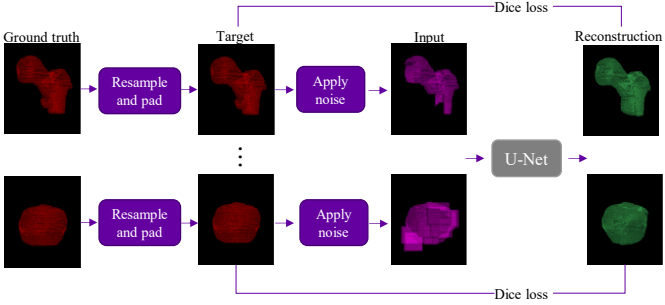

**Fig. 1.** Training of the denoising autoencoder in the MR pelvis use case that includes seven organs. For each organ, we applied noise to the ground truth segmentation after preprocessing. 3D U-Net was trained to denoise the corrupted masks.

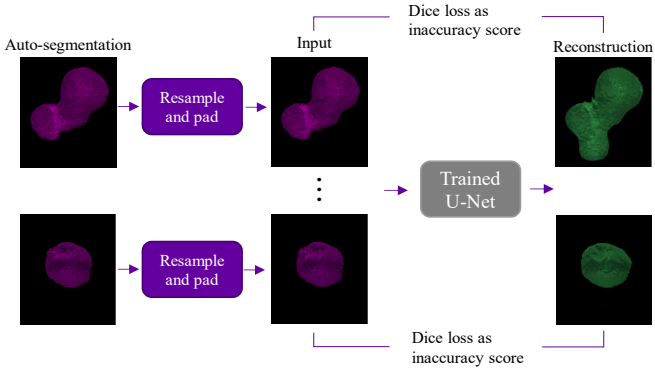

**Fig. 2**. Inference of the denoising autoencoder to detect inaccurate organ auto-segmentations in the MR pelvis use case. For each organ, inaccuracy score was defined as the Dice loss between the preprocessed auto-segmentation and reconstruction.

### 2.5. Comparison Methods

To benchmark our method against existing solutions, we first implemented a VAE-based approach as described in [6]. For each organ, we trained a separate VAE on ground truth segmentations (VAE-Single). We utilized the MONAI framework and selected the hyperparameters to closely match the original method in [6], as no code implementation was available. Furthermore, we trained a multi-organ VAE (VAE-Multi) for each use case, utilizing a larger latent space and increased network depth, hyperparameters are described in Table 2. We were unable to train a deeper VAE-Multi due to instabilities, consistent with findings from previous studies [14], [15]. The preprocessing, loss function, optimizer and model checkpointing approach were the same as for DAE.

In addition, we implemented a statistical approach following [2], [5]. Hui et al. [2] calculated 25 features including volume, surface area, surface area to volume ratio, mean CT number, eccentricity estimates describing the shape, as well as relative features such as displacement vectors between pairs of organ segmentations. The distribution of each feature was parameterized to a best-fit distribution, and outliers were detected based on combinations of deviant features. Altman et al. [5] utilized similar set of features along with intensity-based metrics. Inaccurate segmentations were identified when features exceeded predefined windows determined by feature standard deviations on the development data.

Similar to the other methods, we implemented the statistical approach to be independent of the imaging modality. We utilized volume, surface area, surface area to volume ratio, elongation, roundness, and distance from organ centroid to center defined by the other organs. All the features were calculated using SimpleITK 2.2.1. For each use case, we estimated the mean $\hat{\mu}$ and covariance matrix $\hat{\Sigma}$ of the six features using both the training and validation sets. For a test example with feature vector $x$, we defined the inaccuracy score as the Mahalanobis distance between the example and the distribution with the estimated parameters, i.e., $\sqrt{(x-\hat{\mu})^T \hat{\Sigma}^{-1} (x-\hat{\mu})}$ [16].

| Model | Hyperparameter | Value |
|---|---|---|
| DAE | in_channels, out_channels | 7 (MR pelvis), 2 (CT kidney) |
| | channels | (8, 16, 32, 64, 128, 256, 512, 1024, 2048) |
| | strides | (2, 2, 2, 2, 1, 1, 1, 1) |
| | num_res_units | 2 |
| VAE-Single | out_channels | 1 |
| | latent_size | 10 |
| | channels | (32, 32, 64, 64) |
| | strides | (2, 2, 2, 2) |
| VAE-Multi | out_channels | 7 (MR pelvis), 2 (CT kidney) |
| | latent_size | 100 |
| | channels | (32, 32, 64, 64, 128, 128) |
| | strides | (2, 2, 2, 2, 1, 1) |

**Table 2.** Hyperparameters of the three networks. We used the *UNet* network class for DAE and *VarAutoEncoder* class for VAE-Single and VAE-Multi in *MONAI*.

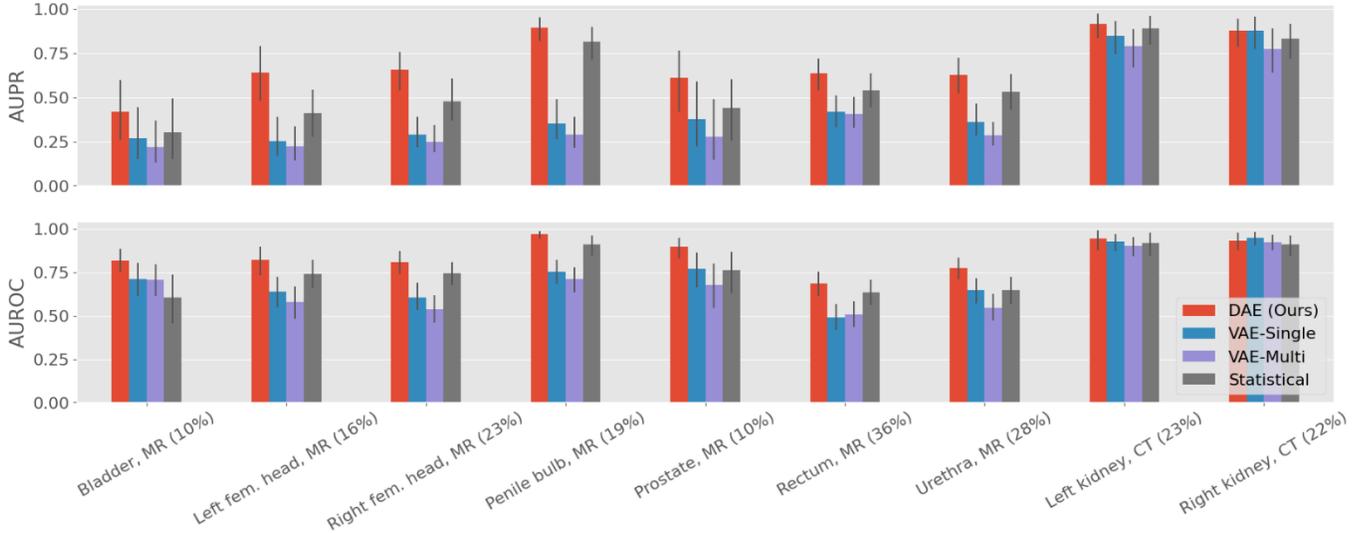

**Fig. 3.** Performance of the four methods in detecting inaccurate organ segmentations, based on AUROC and AUPR. Error bars represent the 95% bootstrap confidence intervals. Percentage of inaccurate segmentations is displayed next to each organ label.

## 3. RESULTS

### 3.1. MR Pelvis

We evaluated the four methods on the auto-segmentations in the test set. Following [4], auto-segmentations were categorized into accurate and inaccurate classes based on the performance of the MR pelvis segmentation models. An organ auto-segmentation was labeled as inaccurate (label 1) if its Dice coefficient with the ground truth was below an organ-specific threshold. The thresholds were 0.86 for bladder, 0.92 for left and right femoral heads, 0.51 for penile bulb, 0.70 for prostate, 0.78 for rectum and 0.28 for urethra. The acceptable Dice coefficient for segmentations varies significantly with organ size [17]. Figure 3 demonstrates the performance of the four methods based on the area under the receiver operating characteristic curve (AUROC) and the area under the precision-recall curve (AUPR). Error bars depict the 95% bootstrap confidence intervals. We can observe that our method shows superior performance compared to the other three approaches for all the seven organs in the MR pelvis use case.

### 3.2. CT Kidney

The thresholds used to categorize the test set auto-segmentations into accurate and inaccurate classes were established based on the performance of the CT kidney segmentation models. For each kidney, an auto-segmentation was labeled as inaccurate if its Dice coefficient with the ground truth was below 0.93. Figure 3 shows that the four methods yield comparable performance for the two kidneys. Our approach achieves the highest performance for left kidney and ranks second for the right kidney, following VAE-Single.

## 4. CONCLUSION

In this study, we developed an algorithm to detect inaccurate organ segmentations utilizing denoising autoencoders. We demonstrated that our method provides superior performance compared to existing solutions in the literature. Through the application of our method to the MR pelvis and CT kidney use cases, we showed that it is independent of the imaging modality. In addition, our solution can be applied post hoc to any segmentation methods without requiring access to the underlying segmentation algorithm. By providing reconstructed segmentations, our method offers visual information about inaccurate regions of the organ segmentations, improving the explainability of the quality assurance system (Figure 4). We envision our solution to be used as a post-processing step in automated treatment planning workflows. By offering explainable quality assurance, our method may foster clinician-AI interactions, thereby enhancing trust in AI solutions in clinical settings.

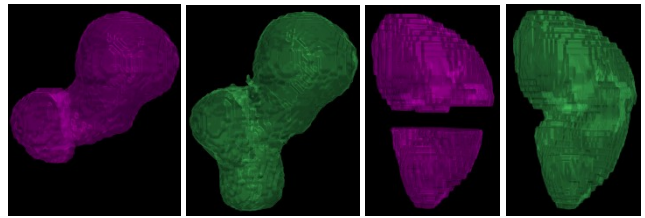

**Fig. 4.** Auto-segmentations (purple) and reconstructed segmentations (green) produced by our method for left femoral head (left two images) and left kidney (right two images). Our method assessed the left femoral head auto-segmentation as under-segmented at the bottom (due to insufficient scan coverage), and the left kidney auto-segmentation as disconnected in the middle (due to malignant tissue eroding the kidney parenchyma).

## 5. COMPLIANCE WITH ETHICAL STANDARDS

For the MR pelvis use case, all the scans were collected privately with the consent of the subjects. Ethical statements can be found in Section 2.8 in Czipczer et al. [7].
CT scans originated from two data sources. 20 cases in the test set were obtained from the publicly available CT-ORG dataset [11], for which ethical approval was not required. The other scans were collected privately from the Chang Gung Memorial Hospital in Taiwan under an approved Institutional Review Board protocol (201701532B0).

## 6. ACKNOWLEDGMENTS

This work was supported by the National Research, Development and Innovation Office (2023-1.1.1-PIACI_FÓKUSZ-2024-00027) and GE HealthCare, where all authors are employed. We thank the annotation team, including authors Zs.K., B.D.-K.; and contributors Kata Bárány, Annamária Cseh, Alinka Csertő, Dóra Dani, Barbara Darázs, Kamilla Dávid, Marcell Dömötör, Janka Fejes, Júlia Hegedűs, Gergely Horváth, Blanka Helga Irmai, Andor Kenyeres, Lotti Lőczi, Eszter Ruff, Krisztina Szepes, Bulcsú Tass, Petra Varga, András Vigh.